\documentclass[english,letterpaper,twocolumn,showpacs,pra,aps]{revtex4}
\usepackage{graphicx}
\usepackage{amssymb}

\makeatletter

\large  

\input epsf

\makeatother

\usepackage{babel}
\makeatother
\begin{document}

\title{Casimir effect due to a single boundary as a manifestation of the Weyl problem}

\author{Eugene B. Kolomeisky$^{1}$, Joseph P. Straley$^{2}$, Luke S. Langsjoen$^{1}$ and Hussain Zaidi$^{1}$}

\affiliation{$^{1}$Department of Physics, University of Virginia, P. O. Box 400714,
Charlottesville, Virginia 22904-4714, USA\\
$^{{2}}$Department of Physics and Astronomy, University of Kentucky,
Lexington, Kentucky 40506-0055, USA}

\begin{abstract}
The Casimir self-energy of a boundary is ultraviolet-divergent. In many cases the divergences can be eliminated by methods such as zeta-function regularization or through physical arguments (ultraviolet transparency of the boundary would provide a cutoff). Using the example of a massless scalar field theory with a single Dirichlet boundary we explore the relationship between such approaches, with the goal of better understanding the origin of the divergences. We are guided by the insight due to Dowker and Kennedy (1978) and Deutsch and Candelas (1979), that the divergences represent measurable effects that can be interpreted with the aid of the theory of the asymptotic distribution of eigenvalues of the Laplacian discussed by Weyl. In many cases the Casimir self-energy is the sum of cutoff-dependent (Weyl) terms having geometrical origin, and an "intrinsic" term that is independent of the cutoff. The Weyl terms make a measurable contribution to the physical situation even when regularization methods succeed in isolating the intrinsic part. Regularization methods fail when the Weyl terms and intrinsic parts of the Casimir effect cannot be clearly separated. Specifically, we demonstrate that the Casimir self-energy of a smooth boundary in two dimensions is a sum of two Weyl terms (exhibiting quadratic and logarithmic cutoff dependence), a geometrical term that is independent of cutoff, and a non-geometrical intrinsic term. As by-products we resolve the puzzle of the divergent Casimir force on a ring and correct the sign of the coefficient of linear tension of the Dirichlet line predicted in earlier treatments. 

\end{abstract}


\maketitle

\section{Introduction}

Casimir interactions are the macroscopic response of the physical vacuum to the introduction of boundaries. They were first derived as an attractive force between perfectly conductive parallel plates induced by the zero-point motion of 
the electromagnetic field \cite{Casimir}.  There is convincing experimental evidence for the reality of these forces \cite{experiment} and a vast body of literature dedicated to various aspects of the phenomenon \cite{CasReviews}. 

The Casimir interaction $\mathcal{E}$ is the difference between the vacuum 
energy of the system constrained by the boundaries and 
that of free space. Since boundaries made of real materials are transparent to sufficiently high-energy modes, the high-energy spectrum is unaffected by the geometry of the system, and only a finite range of the spectrum needs be considered \cite{Casimir,Power}. However, 
in the theoretical treatments of this effect the vacuum energies are usually calculated from an effective low-energy harmonic field 
theory (such as quantum electrodynamics in the case of the electromagnetic Casimir effect), so that
they are approximated by the sum of zero-point energies of a collection of simple harmonic 
oscillators with a spectrum $\omega = c |\textbf{k}|$ (where $c$ is the speed of 
light). In this model, the dispersion relation holds for arbitrarily large wave vectors $\textbf{k}$; both the 
"constrained" and "free" vacuum energy densities are ultraviolet divergent, and 
the Casimir interaction is the difference between two infinite quantities. 
This problem can be resolved by a soft-cutoff modification of the large-\textbf{k} part of the spectrum that leads to a finite vacuum energy. When the result is not sensitive to the form of the cutoff, the Casimir interaction can be extracted by taking the cutoff to infinity at the end of calculation. However, it is more common that there are contributions that diverge in the absence of a cutoff; we will refer to these as "formally divergent terms," since on physical grounds the Casimir energy must be finite. Some approaches to the calculation make use of analytic continuation \cite{zeta} or dimensional regularization \cite{dim} to eliminate the formally divergent terms and extract the "intrinsic" part that is independent of the cutoff. 

The important virtue of the model is that for many geometries the various approaches give the same result for the intrinsic part of the Casimir energy, that is independent of the cutoff and depends only on Planck's constant $\hbar$, the speed of light $c$, and macroscopic length scales. However, this is not always the case: in particular the divergences occurring for spherical geometry in even space dimensions are not removed by any regularization scheme \cite{Bender,Milton}. What this means physically represents an open problem; it seems to imply that in two dimensions a conducting ring placed in vacuum is subject to an infinite stress.  The two-dimensional version of this problem (a circle) was first  studied by Sen \cite{Sen}, who concluded that in addition to the expected cutoff-independent terms, the Casimir energy contains contributions that depend on the  cutoff.  He further found that the formally divergent contributions can be interpreted geometrically.

The main goal of this paper is to analyze the formal divergences encountered in the Casimir self-energy of an object with curved boundary. This is done by looking at the case of a massless scalar field theory with a Dirichlet boundary and explicitly taking into consideration the invisibility of the boundary to sufficiently high-energy field modes. Specifically we are guided by observation due to Dowker and Kennedy \cite{DK} and Deutsch and Candelas \cite{CD} that the formally divergent terms correspond to measurable effects, and that they can be understood and interpreted geometrically through the connection of the Casimir problem to the Weyl problem of the asymptotic distribution of the eigenvalues of the Laplacian \cite{Weyl}. This connection will be illustrated for several geometries by simultaneous solution of the Casimir and Weyl problems and showing that the Weyl piece of the Casimir effect corresponds to the formally divergent contributions. The cases for which it has not been possible to regularize the divergence are shown to belong to a separate class where the Weyl piece of the Casimir energy contains a contribution that is logarithmically dependent on the cutoff. We argue that this is what happens in even space dimensions and demonstrate our contention by detailed analysis of the Casimir effect due to a smooth Dirichlet boundary in two spatial dimensions. As by-products we resolve the problem of the divergent Casimir force exerted on the ring in two dimensions \cite {Bender,Milton} and correct the sign of the coefficient of linear tension given previously \cite{Sen}. 

The organization of this paper is as follows. In Section II we introduce a new computational method in which the Casimir effect is viewed as due to the vacuum fluctuations \textit{eliminated} by introduction of the Dirichlet boundaries. 

In Section III we review the relationship between the Weyl problem of asymptotic distribution of the eigenvalues of the Laplacian and the Casimir problem. As a motivation we calculate the formally divergent Casimir energy of a single Dirichlet plane (Section IIIA) and show how the latter can be understood geometrically in terms of the areal density of states (DOS) of the Weyl problem (Section IIIB). For a field confined to a region the Weyl DOS is defined as a smooth part of the exact DOS. Since the Weyl DOS contains ultraviolet spectral information, it is natural to conjecture that the Weyl DOS is entirely responsible for the formally divergent parts of the Casimir effect. This leads to a large wave vector expansion of the Weyl DOS. The Weyl-Casimir correspondence allows us to deduce the two leading terms of this expansion known as \textit{Weyl's conjecture}. These ideas are illustrated in Section IIIC where we compute the Casimir energies of a field confined to a one-dimensional interval with Dirichlet and periodic boundary conditions. Additionally we demonstrate that the Weyl piece of the Casimir effect is insensitive to the topology of the problem (whereas the intrinsic part is sensitive to this). The analysis of the same problem via the zeta-function regularization technique carried out in Section IIID reveals that the Weyl piece of the effect is usually discarded (but possibly can be retrieved). One of the consequences of the separability of the Weyl and intrinsic parts of the Casimir effect discussed in Section IIIE is an expression for the Weyl energy of the boundary as a surface integral of an even function of curvature invariants. For the special case of a spherical shell in general space dimension, this expression becomes a finite series in powers of the sphere radius. Dimensional analysis of this series reveals the special role of even space dimensions as being cases where there is not a clear separation between the Weyl and cutoff-independent pieces of the Casimir energy. 

In Section IV we compute the Casimir energy in two spatial dimensions due to an arbitrary smooth Dirichlet boundary; this contains the main results of our work. This is the case where regularization techniques fail to give a finite value for the Casimir energy. As an illustration of this statement we show why the zeta-function regularization approach cannot be successfully employed here. Consistent with the observations made in Section IIIE, we demonstrate that even though the Weyl-Casimir correspondence continues to play a prominent role, we no longer have a clear cut separation of the cutoff-dependent and intrinsic parts of the Casimir effect. Our result is an expression for the Casimir energy due to a smooth boundary, in the form of a sum of formally divergent terms coming from the Weyl expansion, a geometrical term that is cutoff independent, and a non-geometrical intrinsic contribution. Apart from the last term that cannot be given in general form, the bulk of the effect has the form of a purely geometrical expression. The latter is deduced from analysis of a particular case (a wedge bounded by an arc) for the Dirichlet and periodic boundary conditions (Sections IVA and IVB, respectively) that can be generalized to the case of an arbitrary curve. In Section IVC these results are applied to compute the cutoff-independent difference in the Casimir energies between the Dirichlet and periodic geometries. 

Our conclusions are summarized in Section V. For all the cases considered in this work we correctly predict the sign of the effect, which is possible because we calculate the Casimir energy in terms of the field modes eliminated by the Dirichlet boundary. Prediction of the sign of the Casimir effect is an important fundamental problem whose general solution is yet to be found. The only rigorous result in this area is a proof \cite{KK} that the Casimir force between two arbitrarily shaped bodies related by reflection is always attractive. Additionally, in the scalar case the sign of the effect can be estimated from the sign of the contribution due to shortest periodic rays \cite{Schaden}. 

\section{Calculating the Casimir energy}

Here we will explain how to calculate the change in energy resulting from the
introduction of a Dirichlet boundary into a quantum system at zero temperature.
For technical reasons, it is useful to first formulate the problem at
finite temperature $T$, and subsequently take the limit $T \rightarrow 0$.
The starting place is a Gaussian field theory having the Euclidean action
\begin{equation}
\label{action}
S_{E}[w] = {\frac{1}{2}}\int_{0}^{\hbar/T} d\tau d^{d}x \left (c^{-2}({\frac{\partial w}{\partial \tau}})^{2} +
(\nabla w)^{2}\right ).
\end{equation}
The real scalar $w$ is a function of the $d$-dimensional position vector $\textbf{r}$ and  is periodic in imaginary time $\tau$, so that $w(\textbf{r},0) = w(\textbf{r}, \hbar/T)$ \cite{Matsubara}. 

The scalar field theory (\ref{action}) can be viewed as a toy version of electricity and magnetism. The divergences encountered for curved geometries exist in both theories, so they are not due to specifics of electricity and magnetism. Therefore we study the problem in the simpler setting of the scalar theory. 

The zero-point energy can be calculated by means of a functional integral that makes use of the correspondence between the Feynman path integral for the $d$-dimensional field theory and the partition function for a $d + 1$-dimensional classical statistical mechanics problem. The path integral is
\begin{equation}
\label{Zdef}
Z_{w} = \int Dw(\textbf{r},\tau) \exp( - S_{E}[w]/\hbar).
\end{equation}
This resembles the partition function for a classical statistical mechanics
problem, with the Hamiltonian replaced by the action $S_{E}$ and the 
temperature replaced by Planck's constant $\hbar$ (please note that this is 
distinct from the real temperature $T$)
\cite{Kogut}.  In this analogy the energy for the quantum
system corresponds to the "free energy" per unit "length" in the imaginary
time direction, so that $\mathcal{E} = \hbar (\ln Z_{w})/(\hbar /T) = - T \ln Z_{w}$, and the zero-point energy is the $T \rightarrow 0$ limit of this.

Now assume that the vacuum is disturbed by the addition of sharp boundaries which constrain the field in some way and thus eliminate some degrees of freedom of the 
vacuum fluctuations. The constrained field (which we will refer to as $v$) inherits 
any boundary conditions imposed on $w$ as well as new conditions on surfaces $D_{i}$ of Dirichlet ($v|_{D_{i}} = 0$) type where the subscript $i$ labels the boundaries. We will write the difference between the original and constrained fields in the form $w(\textbf{r},\tau) = v(\textbf{r},\tau) + u(\textbf{r},\tau)$, where $u(\textbf{r},\tau)$ can be chosen to satisfy the $d+1$-dimensional equation
\begin{equation}
\label{gbvproblem}
(\frac{\partial^{2}}{c^{2}\partial\tau^{2}} + \triangle) u = 0, ~~~~~~u|_{D_{i}} = f_{i}(\textbf{r},\tau)
\end{equation} 
where $f_{i}$ are functions defined on the boundaries and determined by the boundary values of $w$; they play the role of dynamical variables of our approach. The reason for defining $u$ this way is that it eliminates the cross term in the action, so that $S_{E}[w] = S_{E}[v] + S_{E}[u]$. Then the functional integral factors into integrations over $v$ and $u$, so that $Z_{w} = Z_{v} Z_{u}$. Since the zero-point energy for the unconstrained system is determined by $Z_{w}$ 
and the zero-point energy for the constrained system is determined by $Z_{v}$, the Casimir energy is given by $T\rightarrow 0$ limit of $\mathcal{E}= const+T \ln Z_{u}$. This result allows us to calculate directly the change in energy due to the fluctuation modes that have been eliminated. 

A simplification is achieved by expanding all the dynamical variables of the problem into a Fourier series in the imaginary time domain; for example $u(\textbf{r}, \tau) = \sum_{\omega} u_{\omega}(\textbf{r}) \exp i
\omega \tau$
where the Fourier coefficients $u_{\omega}(\textbf{r})$ are solutions to the boundary-value problem for the Helmholtz equation
\begin{equation}
\label{Helmholtz}
(\triangle - \frac{\omega^{2}}{c^{2}})u_{\omega} = 0,~~~~~~ u_{\omega}|_{D_{i}} = f_{\omega,i}(\textbf{r})
\end{equation} 

The calculation of the action $S_{E}[u]$ is further simplified when the identity $(\nabla u)^{2} = div(u\nabla u) - u\triangle u$ is substituted into Eq.(\ref{action}). Then the integral of $div(u\nabla u)$ over $d^{d}x$ transforms into a sum of surface integrals. The remaining integral over $d\tau$ vanishes due to the relation $\triangle u = - \partial^{2} u/c^{2}\partial \tau^{2}$ and the condition of periodicity, $u(\textbf{r},0) = u(\textbf{r}, \hbar/T)$. As a result we find
\begin{eqnarray}
\label{actionsurfaceintegral}
S_{E}[u]& =& \frac{1}{2} \int_{0}^{\hbar/T} d \tau \sum_{i}\int[u\nabla u]_{i}d\textbf{s}_{i} \nonumber\\& = & \frac{\hbar}{2T}\sum_{\omega,i}\int f_{\omega,i}[\nabla u_{-\omega}]_{i} d\textbf{s}_{i}
\end{eqnarray}
Here $[\psi]_{i}$ stands for the discontinuity of $\psi$ across the $i$-th boundary, the summation is performed over all the boundaries and we employed continuity of the field $u=f$ at the Dirichlet boundary. 

Although our approach is applicable to an arbitrary number of the Dirichlet boundaries, in all the cases considered in this paper only one boundary contributes into the integral (\ref{actionsurfaceintegral}). Since we are dealing with a harmonic field theory, the solution to the boundary-value problem (\ref{Helmholtz}) is linear in the surface field $f_{\omega}$ and thus the action $S_{E}$ is a quadratic diagonal form of $f_{\omega}$:
\begin{equation}
\label{form}
S_{E} = \frac{\hbar}{2T} \sum_{\omega,\nu} \frac{|f_{\omega \nu}|^{2}}{\lambda_{\nu}(|\omega|/c)}
\end{equation}
where the subscript $\nu$ (representing one or more indices) labels the normal modes of the field $u$ that have been eliminated by the boundary in question, and $\lambda_{\nu}(|\omega|/c) > 0$ is a set of geometry-dependent coefficients. Then according to the prescription outlined above, the Casimir energy is given by
\begin{eqnarray}
\label{Casenergygeneral}
\mathcal{E}&=&const + \frac{T}{2} \sum_{\omega, \nu} \ln(2\pi T \lambda_{\nu}(|\omega|/c))\nonumber\\&\rightarrow&const +\frac{\hbar}{2\pi}\sum_{\nu}~'\int_{0}^{\infty}d\omega
\ln (2\pi T \lambda_{\nu}(\omega/c))
\end{eqnarray}
where in the second step we took the zero-temperature limit according to the rule $\sum_{\omega} \rightarrow (\hbar/T) \int d\omega/2\pi$. As was already mentioned, any physical boundary is invisible to high-energy modes and from now on this fact is made implicit by adding a prime to all pertinent sums and integrals over normal modes. This means that large energy contributions into the sum in (\ref{Casenergygeneral}) are suppressed by a smooth cutoff function that represents transmission properties of the boundary. We will not need an explicit form of the cutoff function. 

The constant in (\ref{Casenergygeneral}) is such as it cancels the temperature dependence and makes the argument of the logarithm dimensionless. Rigorous determination of this constant requires an analysis of the functional integration measure in (\ref{Zdef}) that led to (\ref{Casenergygeneral}). Instead we provide a heuristic argument that gives the same answer. Since in our approach the Casimir effect arises due to harmonic field oscillators eliminated by the Dirichlet boundary it would suffice to understand the case of one degree of freedom (the harmonic oscillator) and then the result can be generalized to that of the harmonic field theory (\ref{action}).

Let us consider a harmonic oscillator of mass $m$ and natural frequency $\Omega$ whose position is given by the displacement function $x(\tau)$. The Euclidian action 
\begin{eqnarray}
\label{hoaction}
S_{E}[x] &=& \frac{m}{2} \int_{0}^{\hbar/T} d\tau \left ((\frac{dx}{d \tau})^{2} + \Omega^{2}x^{2}\right )\nonumber\\
 &=& \frac{m\hbar}{2T}\sum_{\omega}(\omega^{2}+\Omega^{2})|x(\omega)|^{2}
\end{eqnarray}
is Gaussian and the zero-point energy is given by
\begin{equation}
\label{zpenergy}
\mathcal{E} = const + \frac{\hbar}{2\pi}\int_{0}^{\infty} d\omega \ln\left (\frac{\omega^{2}+\Omega^{2}}{2 \pi T/m}\right )
\end{equation}
We need to determine the additive constant, and also deal with the divergence of the
integral at the upper limit.   Since the physical source of the zero-point energy is the confinement of the
field, Eq.(\ref{zpenergy}) should vanish in the free particle limit  $\omega/\Omega \rightarrow \infty$.
Then subtracting from Eq.(\ref{zpenergy}) its high-frequency limit we obtain
\begin{equation}
\label{ absolutezpenergy}
\mathcal{E} = \frac{\hbar}{2\pi}\int_{0}^{\infty} d\omega \ln\left (\frac{\omega^{2}+\Omega^{2}}{\omega^{2}}\right )\equiv\frac{\hbar \Omega}{2}
\end{equation}
as it should be. A rigorous argument based on careful analysis of the functional integration measure leading from (\ref{zpenergy}) to (\ref{ absolutezpenergy}) has been given by Kleinert \cite{Kleinert}. 

In the Casimir problem the confined field is represented by a collection of harmonic oscillators, and the same rule applied to Eq.(\ref{Casenergygeneral}) gives our final general expression for the Casimir energy: 
\begin{eqnarray}
\label{ absCasimirenergy}
\mathcal{E}&=&\frac{\hbar }{2\pi} \sum_{\nu}~' \int_{0}^{\infty}d\omega\ln \frac{\lambda_{\nu}(\omega/c)}{\lambda_{\nu}(\infty)}\nonumber\\&=& - \sum_{\nu}~' \int_{0}^{\infty}\frac{d\omega}{\pi}\frac{\hbar \omega}{2}\frac{d}{d\omega}\ln \frac{\lambda_{\nu}(\omega/c)}{\lambda_{\nu}(\infty)}
\end{eqnarray}
The infinity in the argument of $\lambda_{\nu}$ is not to be taken literally: in any problem the variable $\omega$ would enter via the dimensionless combination $\omega/c$ times a macroscopic length scale. The infinity means that the limit of this length scale going to infinity is to be taken so that the zero-point energy is the confinement energy relative to a reference free field geometry.   Our prescription (\ref{ absCasimirenergy}) agrees with  that of Brevik and Elizalde \cite{Brevik} employed in their calculation of the Casimir energy of a piecewise uniform string.

The rule (\ref{ absCasimirenergy}) guarantees that the integral over $\omega$ is ultraviolet convergent and that every field oscillator of frequency $\Omega_{\nu}$ eliminated by the boundary contributes $-\hbar \Omega_{\nu}/2$ into the Casimir energy. The second representation of (\ref{ absCasimirenergy}) (where we integrated by parts) hints at a relationship between our approach and the calculation of the Casimir energy by means of a contour integral \cite{contour,Milton}.

We note that the integration variable $\omega$ in Eq.(\ref{ absCasimirenergy}) has its origin in the Fourier transform in the imaginary time domain that led from Eq.(\ref{gbvproblem}) to Eq.(\ref{Helmholtz}). This auxiliary variable is not related to the physical frequency of a field oscillator and therefore it should not be subject to any large-$\omega$ cutoff. On the other hand, the mode index $\nu$ is a physical variable akin to a wave vector and thus it should be subject to a cutoff as a consequence of the penetrability of the boundary to high-energy modes. 

Our cutoff rule is different from that employed by Sen \cite{Sen} in his analysis of the circle geometry. Expressed in our language, he imposes an exponential cutoff on the variable $\omega$ but leaves the mode variable $\nu$ unaffected. Although his procedure gives a finite answer for the circular geometry, this would not be the case for the cylinder and many other geometries. On the other hand, our rule (\ref{ absCasimirenergy}) is universally applicable.

\section{Geometrical interpretation of ultraviolet divergences}

Let us assume that the physical boundary is characterized by a frequency cutoff at scale $\omega_{0}$. The boundary is impenetrable to low-energy field modes but invisible to the modes of energy significantly greater than $\hbar\omega_{0}$. Such a boundary can be modeled by a Dirichlet surface. We will determine the coefficient $\gamma_{0}$ of the fluctuation-induced surface tension of a single Dirichlet plane immersed in a $d$-dimensional vacuum. Since this problem is only characterized by microscopic energy and length scales, $\hbar\omega_{0}$ and $c/\omega_{0}$, respectively, the outcome can be estimated via dimensional analysis. Indeed the only quantity having dimension of energy divided by length to the power $d-1$ is $\gamma_{0}\sim \hbar\omega_{0}/(c/\omega_{0})^{d-1}=\hbar c(\omega_{0}/c)^{d}$. We observe that the surface tension diverges as the cutoff frequency $\omega_{0}$ approaches infinity. This divergence has no effect on the Casimir stress on parallel planes because the overall area of the boundaries remains fixed as the distance between the planes is varied. 

However this is not the case for curved boundaries. For example, a change of the radius of a circle implies a change of the perimeter and as a result the formally divergent self-energy will contribute to the Casimir force per length. This idea, originally due to Deutsch and Candelas \cite{CD}, was recently re-expressed by Graham, Jaffe and co-workers \cite{Jaffe} and by Barton \cite{Barton}. The implication is that in an experimental situation, a curved boundary might give rise to a large cutoff-dependent contribution
to the physically measurable Casimir force \cite{critique} (if the boundary is not strictly rigid). This is consistent with Sen's observation \cite{Sen} that the coefficients of the cutoff-dependent terms contributing to the Casimir energy are geometrical objects such as the length of the circular boundary. We also note that in the high-temperature limit the Casimir free energy is expressed in terms of a surface integral of a quadratic function of local curvature \cite{BD} which can be shown to have entirely geometrical nature.

In order to better understand the geometrical meaning of the surface divergence we need to explicitly compute the Casimir energy due to a Dirichlet plane. This can be done by the method outlined in the previous Section.

\subsection{Surface energy of a plane in $d$ dimensions}

Let us consider single Dirichlet plane perpendicular to the $z$-axis of the $d$-dimensional rectangular coordinate system. Since the space is uniform relative to translations parallel to the boundary, both the fields $u_{\omega}(\textbf{r})$ and $f_{\omega}(\textbf{r}_{\perp})$ (where $\textbf{r}_{\perp}$ is the position vector perpendicular to the $z$ axis) can be expanded into a Fourier series; for example $u_{\omega}(\textbf{r}) = \sum_{\small\textbf{q}} u_{\omega \small\textbf{q}}(z) \exp i \textbf{q} \textbf{r}_{\perp}$. The boundary-value problem (\ref{Helmholtz}) for the Fourier coefficients $u_{\omega \small\textbf{q}}(z)$ becomes
\begin{equation}
\label{bvplane}
(\frac{d^{2}}{dz^{2}} - q^{2} - \frac{\omega^{2}}{c^{2}})u_{\omega \small\textbf{q}}(z) = 0, ~u_{\omega \small\textbf{q}}(0) = f_{\omega \small\textbf{q}}
\end{equation}
where we assumed that the plane is located at $z=0$. The particular solution (i.e. vanishing for $f=0$) to (\ref{bvplane}) that decays as $|z|\rightarrow \infty$ is
\begin{equation}
\label{planesolution}
u_{\omega \small\textbf{q}}(z) = f_{\omega \small\textbf{q}} \exp \left (- \sqrt{q^{2}+\omega^{2}/c^{2}}|z| \right )
\end{equation}
Substituting the Fourier representations for $u_{\omega}(\textbf{r})$ and $f_{\omega}(\textbf{r}_{\perp})$ along with the solution (\ref{planesolution}) in Eq.(\ref{actionsurfaceintegral}) we find
 \begin{equation}
\label{actionplanar}
S_{E} = \frac{\hbar}{2T} \sum_{\omega, \small\textbf{q}}2\mathcal{A}\sqrt{q^{2}+\omega^{2}/c^{2}}|f_{\omega \small\textbf{q}}|^{2}
\end{equation}
where $\mathcal{A}$ is the macroscopic $(d - 1)$-dimensional area of the 
boundary. The action (\ref{actionplanar}) has the required form Eq.(\ref{form}), with the geometrical coefficient $\lambda_{\textbf{q}}(\omega/c)= 1/(2\mathcal{A}\sqrt{q^{2}+\omega^{2}/c^{2}})$ 
that becomes small for large $\textbf{q}$. According to Eq.(\ref{planesolution}) the disturbance
introduced by the fluctuating boundary condition is localized at the
boundary, to within a length that is proportional to $\lambda$ itself. This is the source of the divergent coefficient of the surface tension. Substituting this $\lambda$ in Eq.(\ref{ absCasimirenergy}) and integrating over $\omega$ we find
\begin{equation}
\label{surfaceenergysum}
\mathcal{E}=-\frac{\hbar}{4\pi}\sum_{\textbf{q}}~'\int_{0}^{\infty}d\omega \ln\frac{\omega^{2}+c^{2}q^{2}}{\omega^{2}}= - \frac{1}{2} \sum_{\textbf{q}}~'\frac{\hbar cq}{2}
\end{equation}
The sum in (\ref{surfaceenergysum}) can be recognized as the zero-point energy of a harmonic field confined to a $d-1$-dimensional space. By eliminating a range of the field modes, the Dirichlet boundary suppresses vacuum fluctuations in a region near it. This lowers the energy relative to the freely fluctuating vacuum, with the result that 
the Casimir energy due to the Dirichlet boundary (\ref{surfaceenergysum}) is negative. The magnitude of the effect is a fraction of the $d-1$-dimensional zero-point energy because the surface energy is dominated by the field modes highly localized at the boundary (without the cutoff this surface energy would be ultraviolet divergent).

For macroscopic area $\mathcal{A}$ of the boundary the sum in (\ref{surfaceenergysum}) can be transformed into an integral with the result
\begin{equation}
\label{surfaceenergyintegral}
\mathcal{E}=-\frac{\hbar c \mathcal{A}K_{d-1}}{4}\int_{0}^{\infty~'}q^{d-1}dq
\end{equation}
where in taking the macroscopic limit we used the rule $\sum_{\small\textbf{q}} \rightarrow \mathcal{A} \int d^{d-1}q/(2\pi)^{d-1}$. The parameter $K_{d}$ is the surface area of a $d$-dimensional unit sphere, $S_{d}=2\pi^{d/2}/\Gamma(d/2)$, divided by $(2\pi)^{d}$:
\begin{equation}
\label{KD }
K_{d}=\frac{2\pi^{d/2}}{(2\pi)^{d}\Gamma(d/2)}
\end{equation}
We note that for $d>1$ the coefficient of surface tension is negative and given by $\gamma_{0}=\mathcal{E}/\mathcal{A}\sim-\hbar c (\omega_{0}/c)^{d}$ which is in accordance with the dimensional argument given earlier. For a smooth curved Dirichlet boundary, Eq.(\ref{surfaceenergyintegral}) represents the leading term of a geometric expansion to be discussed in more detail below. For $d=1$ the surface energy (\ref{surfaceenergyintegral}) vanishes. This circumstance does not contradict our argument that introduction of the Dirichlet surface lowers the vacuum energy. It just means that in one dimension the leading finite-size correction to the bulk vacuum energy is negative, cutoff-independent, and of the order $\hbar c$ divided by the system size. This correction term is the difference between the sum (\ref{surfaceenergysum}) and the integral (\ref{surfaceenergyintegral}); an explicit demonstration of this fact will be given below.

\subsection{Weyl expansion and the formally divergent part of the Casimir effect}

Although the surface energy is cutoff-dependent and determined by the transmission properties of the boundary, there is an intrinsic feature built into (\ref{surfaceenergyintegral}) that can be revealed by rewriting the expression for the surface energy in a form that makes connection to its origin as a zero-point effect:
\begin{equation}
\label{ surfaceenergydosform}
\mathcal{E}= \int_{0}^{\infty~'}\frac{\hbar cq}{2} g_{area}(q)dq
\end{equation}
where
\begin{equation}
\label{arealdos}
g_{area}(q)=-\frac{1}{2}\mathcal{A}K_{d-1}q^{d-2}
\end{equation}
can be called the areal density of states; it is a geometrical (and therefore generally relevant) quantity. This is an instance of a common feature, apparently first recognized by Dowker and Kennedy \cite{DK} and by Deutsch and Candelas \cite{CD}, and recently emphasized by Schaden \cite{Schaden} that \textit{all} cutoff-dependent contributions into the Casimir effect have a geometrical nature interpretable in terms of the DOS. Specifically, in the case of the scalar Casimir effect considered in this paper, they all can be understood as having their origin in the asymptotic limit of the density of eigenvalues of the Laplacian. The relationship to this classic problem of mathematical physics posed by Lorentz, tackled by Weyl and other researchers \cite{Weyl} and made popular by Kac \cite{Kac} can be summarized as follows: 

The zero-point energy of the field confined to a region is the sum of zero-point energies of the field oscillators 
\begin{equation}
\label{zpstandard }
\mathcal{E} = \sum_{\nu}~'\frac{\hbar cq_{\nu}}{2}\equiv\int_{0}^{\infty~'}\frac{\hbar cq}{2} G(q)dq
\end{equation}
where $-q_{\nu}^{2}$ are eigenvalues of the Laplacian:
\begin{equation}
\label{Laplacian}
(\triangle + q^{2})w=0
\end{equation}
The discrete eigenvalue spectrum is determined by imposing the Dirichlet boundary condition $w(\textbf{r})=0$ at the boundary of the region. The function
\begin{equation}
\label{exactDOS}
G(q) = \sum_{\nu}\delta(q-q_{\nu})
\end{equation}
is the exact DOS. It can be represented as a histogram with the distance between neighboring peaks of the order of the inverse of the system size. Quite generally \cite{Weyl} the exact density of eigenvalues can be presented as a sum of the smooth (semi-classical) part $g(q)$, commonly referred to as the Weyl DOS, and an oscillatory remainder $G(q)-g(q)$. The Weyl DOS $g(q)$ can be understood as the result of averaging of the exact DOS (\ref{exactDOS}) over scales in $q$-space that exceed the distance between the neighboring peaks of $G(q)$ \cite{Weyl}. Therefore $g(q)$ contains information about the large-$q$ behavior of the exact DOS, and the cutoff-dependent part of the Casimir energy can be explained as having its origin in the Weyl DOS. Likewise, the remainder $G(q)-g(q)$ is responsible for the intrinsic part of the zero-point energy. It is then expected that in the macroscopic limit the zero-point energy (\ref{zpstandard }) naturally splits into cutoff-dependent and intrinsic pieces:
\begin{equation}
\label{splitzpenergy }
\mathcal{E}=\int_{0}^{\infty~'}\frac{\hbar cq}{2} g(q)dq+\int_{0}^{\infty}\frac{\hbar cq}{2}[G(q)-g(q)]dq
\end{equation}
In view of its ultraviolet (local) origin, the first cutoff-dependent term in (\ref{splitzpenergy }) is \textit{additive}, i.e. it can be presented as a sum of volume, surface, line etc. boundary integrals. This is however not the case for the second (cutoff-independent) term where we have taken the limit of infinite cutoff frequency; requiring that the integral converges is a part of the definition of the Weyl DOS $g(q)$.

The smooth part of the exact DOS, $g(q)$, can be represented as a large-$q$ expansion and each term of this Weyl expansion, similar to Eq.(\ref{arealdos}), can be interpreted geometrically. Indeed, for a $d$-dimensional volume $\mathcal{V}$ enclosed by a $(d-1)$-dimensional Dirichlet boundary of area $\mathcal{A}$ the Weyl expansion starts out as
\begin{equation}
\label{WeylDOS2terms}
g(q)= \mathcal{V}K_{d}q^{d-1}-\frac{1}{4}\mathcal{A}K_{d-1}q^{d-2}+...
\end{equation} 
where the first term is the well-known consequence of the transformation rule 
$\sum_{\textbf{q}}\rightarrow \mathcal{V}\int d^{d}q/(2\pi)^{d}$. 
For an infinite vacuum only the first term is present; when this is the reference state the Casimir
energy can be calculated from 
Eq.(\ref{splitzpenergy }) by removing the leading term from Eq. (\ref{WeylDOS2terms}).
The second term, 
proportional to the area $\mathcal{A}$ of the boundary, only includes the effects of modes inside the region and thus is half as large as
(\ref{arealdos}) 
where the boundary has two sides.
For $d=2,3$ the terms displayed in (\ref{WeylDOS2terms}) are Weyl's 
original results \cite{Weyl,Kac}.   For general $d$ an equivalent expression for the \textit{number} of states
$\int g(q) dq$ has been given by Brownell \cite{Brownell}.

Even though the Weyl DOS is a purely geometrical concept having little to do with physics, its relationship to the Casimir problem explains the sign of the surface term in Eq.(\ref{WeylDOS2terms}). 

A series of examples illustrating the general principles just outlined has been given by Schaden \cite{Schaden}. Below we consider two one-dimensional geometries that illustrate the expectation (\ref{splitzpenergy }). Additionally they contain the ingredients needed in analysis of the scalar Casimir effect in two dimensions where we will find deviations from the rule (\ref{splitzpenergy }).

\subsection{One-dimensional interval}

Consider a harmonic field confined to a one-dimensional Dirichlet interval of length $s$. The eigenvalue spectrum is $q_{n}=\pi n/s$, $n=1, 2, ...$ and the exact DOS is given by 
\begin{equation}
\label{1dexactDOS}
G(q) = \sum_{n=1}^{\infty}\delta(q-\frac{\pi n}{s})
\end{equation}
Since the peaks of this function are equidistant, the Weyl DOS can be readily obtained by noticing that every interval of length $\triangle q = \pi/s$ (that does not include the origin $q=0$) contains exactly one eigenvalue. This implies that $G(q)$ averaged over the interval of length $\triangle q$ gives the Weyl DOS, $g(q)= s/\pi$. This agrees with the general result (\ref{WeylDOS2terms}) and confirms that the surface (edge) term is indeed absent. 

The zero-point energy is given by 
\begin{equation}
\label{1dzpenergy}
\mathcal{E}=\frac{\pi \hbar c}{2s}\sum_{n=1}^{\infty}~'n
\end{equation} 
In the macroscopic limit $\omega_{0}s/c\gg1$ that we are interested in, the sum in (\ref{1dzpenergy}) can be computed with desired accuracy with the help of the Euler-Maclaurin summation formula \cite{integral}
\begin{equation}
\label{EM}
\sum_{n=1}^{\infty}~'F(n)\approx\int_{0}^{\infty~'}F(x)dx - \frac{1}{2}F(0)-\frac{1}{12}F'(0)
\end{equation}
Applying this formula to the sum in Eq.(\ref{1dzpenergy}) and employing $q=\pi n/s$ we find the expression 
\begin{equation}
\label{1dzpenergysplit}
\mathcal{E} = \frac{\pi \hbar c}{2s}\left (\int_{0}^{\infty'}xdx - \frac{1}{12}\right ) = \int_{0}^{\infty'}\frac{\hbar cq}{2}\frac{sdq}{\pi} - \frac{\pi \hbar c}{24s}
\end{equation}
that is in agreement with the general expectation (\ref{splitzpenergy }): the cutoff-dependent piece of the effect proportional to the system size $s$ indeed originates from the Weyl DOS given by the $d=1$ limit of (\ref{WeylDOS2terms}), $g(q)=s/\pi$. The attractive intrinsic part of the zero-point energy, $-\pi \hbar c/(24s)$, is well-known \cite{conformal}; its sign was anticipated at the end of Section IIIA.

The Casimir energy is known to be sensitive to the topology of the problem but this sensitivity is limited to the non-additive intrinsic part of the phenomenon because it has non-local origin. As a simple illustration of this principle, let us consider a harmonic field confined to a one-dimensional interval of length $s$ with periodic boundary conditions. The eigenvalue spectrum is $q_{n}=2\pi n/s$, $n=0, \pm 1, \pm2, ...$ and the zero-point energy is given by 
\begin{eqnarray}
\label{1dzpenergyperiodic }
\mathcal{E}=\frac{2\pi \hbar c}{s}\sum_{n=1}^{\infty}~'n&\rightarrow& \frac{2\pi \hbar c}{s}\left (\int_{0}^{\infty'}xdx-\frac{1}{12}\right )\nonumber\\
&=&\int_{0}^{\infty'}\frac{\hbar cq}{2}\frac{sdq}{\pi}-\frac{\pi \hbar c}{6s}
\end{eqnarray}
where we applied the Euler-Maclaurin formula (\ref{EM}) and employed $q=2\pi n/s$. We observe that the first cutoff-dependent terms of (\ref{1dzpenergysplit}) and (\ref{1dzpenergyperiodic }) are identical: both cases are described by the same Weyl DOS, and the double degeneracy of the $|n|>0$ eigenvalues in periodic geometry is compensated, compared to the Dirichlet case, by twice the distance between nearest $q$'s. Only the last cutoff-independent terms of (\ref{1dzpenergysplit}) and (\ref{1dzpenergyperiodic }) sense the difference in topology of the problems. The factor of four difference between the intrinsic parts of (\ref{1dzpenergysplit}) and (\ref{1dzpenergyperiodic }) is well-known \cite{conformal}. The negative sign of the last term of (\ref{1dzpenergyperiodic }) can be understood by realizing that the modes eliminated by the periodic geometry do not contribute into the zero-point energy.

\subsection{Relationship to the zeta-regularized result}

At this point it is instructive to see how approach based on the concept of the Weyl DOS is related to regularization approaches that do not use physical cutoffs. To be specific, let us again consider the one-dimensional Dirichlet interval of length $s$. Then the ultraviolet divergence present in the cutoff-free energy sum (\ref{1dzpenergy}) can be interpreted in terms of the Riemann $\zeta$-function \cite{zeta} 
\begin{equation}
\label{zeta}
\zeta(\sigma) = \sum_{n=1}^{\infty}n^{-\sigma}
\end{equation} 
which is convergent for $\sigma > 1$ and can be analytically continued to all complex $\sigma \neq 1$. Then Eq.(\ref{1dzpenergy}) is the value at $\sigma = -1$ of the expression 
\begin{equation}
\label{1d_zeta}
\mathcal{E}^{(R)}(\sigma)= \frac{\pi \hbar c}{2s}\zeta(\sigma) 
\end{equation}
where the superscript $R$ stands for "regularized." Using the result 
$\zeta(-1) = -1/12$, this gives 
\begin{equation}
\label{1d_zeta2}
\mathcal{E}^{(R)}= \frac{\pi \hbar c}{2s}\zeta(-1)= - \frac{\pi \hbar c}{24s}
\end{equation}
which is the intrinsic part of (\ref{1dzpenergysplit}). The cutoff-dependent term of Eq.(\ref{1dzpenergysplit}) that is linear in $s$ seems to have been discarded. However, we observe that it is possible to retrieve the cutoff-dependent term from the zeta-regularized result as follows: the expression (\ref{1d_zeta}) has a pole at $\sigma = 1$, which is between the range $\sigma > 1$ where the sum (\ref{1d_zeta}) is convergent and the value $\sigma = -1$ where we wish to evaluate (\ref{1d_zeta}).  We believe that in general  there is a cutoff-dependent term associated with every such "bypassed pole"; when the zeta-regularized expression is to be evaluated at $\sigma = -1$ and has a pole at $\sigma = m$ with residue $A$, the corresponding cutoff-dependent term has form $\int 'q^{m} dq$ with a coefficient that is proportional to $A$.  Then the 
zeta-function regularization method would give an elegant route to the determination of all of the terms in the Casimir energy.  In fact Elizalde demonstrated that progress in this direction can be made if the zeta-function technique is supplemented with the Hadamard regularization method 
\cite{Elizalde}.

\subsection{Weyl energy and geometry of the boundary}

A natural consequence of the separability of the Weyl and intrinsic pieces of the Casimir energy is the conclusion that the Weyl energy $\mathcal{E}(\omega_{0})$ given by the first term of (\ref{splitzpenergy }) can be expressed in the form of a purely geometrical expression. When the media on both sides of the boundary have the same speed of light (so that locally the boundary cannot distinguish the inside from the outside), the Weyl energy is given by a surface integral of an "even" combination of curvature invariants that does not depend on the sense of the local normal (the contributions from the "odd" terms cancel). For example, for a smooth boundary in three dimensions $\mathcal{E}(\omega_{0})$ has the form \cite{CD}
\begin{equation}
\label{membrane}
\mathcal{E}(\omega_{0}) = \int ds (\gamma_{0} + \gamma_{1a}(\mathcal{C}_{1}-\mathcal{C}_{2})^{2}+\gamma_{1b}\mathcal{C}_{1}\mathcal{C}_{2})
\end{equation}
where $\gamma_{0}$ is the coefficient of surface tension mentioned earlier while $\gamma_{1a,b}$ are the curvature stiffness constants and $\mathcal{C}_{1,2}$ are the principal curvatures. Since the boundary is made of real material, the shape coefficients $\gamma_n$ have to be interpreted as contributions to the elastic constants of the boundary viewed as a flexible membrane. Assuming locality, the energy expression (\ref{membrane}) can be written down phenomenologically without referring to the Weyl problem. The latter however puts (\ref{membrane}) on a solid footing. Like the coefficient of surface tension $\gamma_{0}\sim \hbar \omega_{0}^{3}/c^{2}$, the curvature stiffness constants $\gamma_{1a,b}$ are dominated by highly localized field components that are suppressed by the boundary. Thus they can only depend on the microscopic energy $\hbar \omega_{0}$ and the length scale $c/\omega_{0}$; they are properties of the material of which the boundary is made, and are the same for all objects of whatever shape made from this material. Dimensional analysis then implies that $\gamma_{1a,b} \sim \hbar \omega_{0}$ \cite{CD}. We now see that the Weyl energy (\ref{membrane}) is an expansion in powers of the cutoff frequency $\omega_{0}$ with the leading $\omega_{0}^{3}$ term and sub-leading $\omega_{0}$ term. In the macroscopic limit that we consider, there is no need to take into account powers of curvature invariants higher than those included in Eq.(\ref{membrane}) because the corresponding curvature constants vanish in the $\omega_{0}\rightarrow \infty$ limit. We also note that the geometrical expression (\ref{membrane}) is applicable to an arbitrary harmonic theory and arbitrary boundary conditions; however, the precise values of the stiffness constants $\gamma$'s do depend on the nature of the field and type of boundary conditions.

 For simple geometries the expression for the Weyl energy (\ref{membrane}) allows us to draw conclusions regarding whether the Casimir self-stress experienced by the boundary is independent of the cutoff. Indeed, the Casimir energy for a spherical shell of radius $a$ is expected to have the form \cite{CD}
\begin{equation}
\label{sphere}
\mathcal{E}_{sphere} = 4\pi \gamma_{0} a^{2} + 4\pi \gamma_{1b} + \# \frac{\hbar c}{a}
\end{equation} 
where the first two terms follow from (\ref{membrane}) while the functional form of the last term is dictated by dimensional analysis; the numerical prefactor $\#$ is determined by the nature of the field involved and by the boundary conditions the shell imposes on the field. For the case of the electromagnetic field the coefficient of surface tension is known to be zero, $\gamma_{0} = 0$ \cite{Boyer0}. This means that the pressure on the boundary is independent of the cutoff and is correctly determined by regularization methods. At the same time it is clear that such a behavior is an exception rather the rule. Indeed for the scalar Casimir effect with Dirichlet or Neumann boundary conditions the coefficient of surface tension is non-zero, and the Casimir pressure is dominated by the first term of (\ref{sphere}).

For another example let us consider the case of an infinite cylindrical shell of radius $a$. In the limit that the cylinder length $L$ goes to infinity, its Casimir energy per unit length along the cylindrical axis is given by 
\begin{equation}
\label{cylinder}
\frac{\mathcal{E}_{cylinder}}{L} = 2\pi a\gamma_{0}+ \frac{2\pi \gamma_{1a}}{a}+\# \frac{\hbar c}{a^{2}}
\end{equation}
As in the spherical case (\ref{sphere}), the cutoff-dependent terms are determined by Eq.(\ref{membrane}) while the functional form of the last piece is fixed by dimensional analysis. We now see that even in the case of the electromagnetic field, $\gamma_{0} = 0$, the pressure on the shell surface is dominated by the cutoff-dependent $2\pi \gamma_{1a}/a$ contribution.

\subsection{Why are even space dimensions special?}

The analysis conducted up to this point assumed separability of the Weyl and intrinsic pieces of the Casimir effect (see Eq.(\ref{splitzpenergy })). In order to see that this assumption breaks down in even space dimensions, let us consider a sphere of radius $a$ in $d$ spatial dimensions. Assuming separability, the Weyl energy of the spherical shell can be estimated to have the form
\begin{equation}
\label{d_sphere}
\mathcal{E}(\omega_{0})\sim \sum_{n=0}^{M}\gamma_{n}a^{-2n}a^{d-1}=\sum_{n=0}^{M}\gamma_{n}a^{d-1-2n}
\end{equation} 
where in the first representation $a^{-2n}$ stands for the nonvanishing even power curvature terms and $a^{d-1}$ represents the surface area of the boundary; numerical factors have been dropped for clarity. The rigidity constants $\gamma_{n}$ have dimensionality of $energy / length^{d-1-2n}$, and they are determined by the microscopic physics, i. e. by $\hbar \omega_{0}$ and $c/\omega_{0}$, the microscopic energy and length scales, respectively. This implies
\begin{equation}
\label{d_sphere_final}
\mathcal{E}(\omega_{0}) \sim \hbar c \sum_{n=0}^{M}(\frac{\omega_{0}}{c})^{d-2n}a^{d-1-2n}
\end{equation}
We note that the functional form of the expression for the Weyl energy $\mathcal{E}(\omega_{0})$ is more restrictive than what would be implied by a naive dimensional analysis that involves the energy scale $\hbar \omega_{0}$ and the two length scales $c/\omega_{0}$ and $a$. 

The number of formally divergent terms $M+1$ in the expression for the Weyl energy $\mathcal{E}(\omega_{0})$ is fixed by the condition $d-2M \geq 0$. For odd $d$ there are $(d+1)/2$ terms and the least divergent of these is linear in $\omega_{0}$. For even $d$ there are $\frac{d}{2}+1$ terms and the least divergent term is independent of the cutoff $\omega_{0}$. This however contradicts the expectation that the Weyl energy only contains the cutoff-dependent parts of the Casimir effect. The phenomenological resolution of this difficulty lies in allowing for a logarithmic cutoff dependence that replaces $\omega_{0}^{0}$ in the expansion. Indeed if the Weyl DOS expansion contains a term proportional to $1/q^{2}$, then the first term of (\ref{splitzpenergy }) is logarithmically divergent at low energies. Physically the only low-energy cutoff in the problem is set by the inverse system size $1/a$. If this has to be invoked, then locality built into the geometrical expression for the Weyl energy is violated. This implies that when $d$ is even, the Casimir energy has the usual cutoff-dependent (geometrical) and intrinsic (implied by dimensional analysis) contributions, and in addition a contribution of the form
\begin{equation}
\label{nearly}
\mathcal{U}_{even} \sim \frac{\hbar c}{a}\ln\frac{\omega_{0}a}{c}
\end{equation}
with a universal amplitude and logarithmically weak dependence on the cutoff frequency $\omega_{0}$. In the zeta-function regularization approach, this logarithmic
divergence appears as a pole exactly at $\sigma = -1$, where the regularized 
expression is to be evaluated. The detailed analysis of the two-dimensional case conducted below confirms this expectation, and in addition reveals the presence of an extra contribution into the Casimir effect that is both geometrical and independent of the cutoff.

\section{Casimir effect in two spatial dimensions}

The two leading terms displayed in Eq.(\ref{WeylDOS2terms}) are not sufficient to completely understand the cutoff dependent part of the Casimir effect in more than one dimension, and a more accurate expression is needed. An efficient indirect method of generating higher order terms of the Weyl expansion is via the Dirichlet series \cite{Weyl,Kac}
\begin{equation}
\label{ Dirichlet}
K(t)= \sum_{\nu}\exp(-q_{\nu}^{2}t)\equiv\int_{0}^{\infty}\exp(-q^{2}t)G(q)dq
\end{equation} 
whose $t\rightarrow 0$ behavior probes the large-$q$ piece of $G(q)$, i.e. the Weyl DOS $g(q)$. The latter, as implied by (\ref{ Dirichlet}), can be extracted from the inverse Laplace transform of $K(t\rightarrow 0)$ \cite{Weyl,Kac}.

Below we present a detailed analysis of the two-dimensional geometry and determine both the cutoff-dependent and intrinsic parts of the Casimir effect. As a by-product we extract the Weyl DOS $g(q)$ to desired accuracy. Our main result is that in the macroscopic limit the Casimir energy due to a smooth Dirichlet boundary $\Gamma$ is given by
\begin{eqnarray}
\label{main_result }
\mathcal{E}&=&\int_{0}^{\infty~'}\frac{\hbar cq}{2} \left (-\frac{\mathcal{A}dq}{2\pi}\right )\nonumber\\&+&\int_{1/S}^{\infty~'}\frac{\hbar cq}{2}\left (-\frac{dq}{128\pi q^{2}}\int_{\Gamma} \mathcal{C}^{2}(s)ds\right )\nonumber\\&-&\frac{\hbar c \gamma}{256\pi}\int_{\Gamma} \mathcal{C}^{2}(s)ds + \mathcal{U}_{na}
\end{eqnarray} 
where $S$ is a macroscopic length scale specific to the problem in question, $s$ is the length along the boundary measured relative to an arbitrary reference point, $\mathcal{C}(s)$ is the curvature at location $s$, and $\gamma=0.57722$ is Euler's constant. The Weyl DOS associated with (\ref{main_result }) is 
\begin{equation}
\label{ 2dWeylDOs}
g(q)= -\frac{\mathcal{A}}{2\pi}-\frac{1}{128\pi q^{2}}\int_{\Gamma} \mathcal{C}^{2}(s)ds .
\end{equation} 
The first term is the $d=2$ case of the general result (\ref{arealdos}); it is proportional to the total length of the boundary $\mathcal{A}=\int_{\Gamma} ds$. The second term proportional to integrated square of local curvature agrees 
with that inferred from the $t\rightarrow 0$ expansion of Eq.(\ref{ Dirichlet}) due to Stewartson and Waechter \cite{geometry}. 

The third term of (\ref{main_result }) is cutoff-independent, even though it is
geometry-dependent. This is a special feature of two dimensions. The first three terms of (\ref{main_result }) accumulate a part of the Casimir energy 
which is dependent on the geometry of the boundary, of the form
\begin{equation}
\label{integralfinite}
\mathcal{E}_{geom} = \int_{0}^{\infty'}\frac{\hbar cq}{2}\left (-\frac{\mathcal{A}dq}{2\pi}\right ) - \frac{\hbar c \ln \frac{\omega_{0}S}{c}}{256\pi}\int_{\Gamma}\mathcal{C}^{2}(s)ds .
\end{equation}
The integral leading to the second term of (\ref{integralfinite}) is evaluated with logarithmic accuracy and the outcome is combined with the third term of (\ref{main_result }). The presence of the infrared cutoff (set by finite system size) in the marginal second term of (\ref{main_result }) is necessary; it is another deviation from the expectation (\ref{splitzpenergy }) that is due to the two-dimensionality of the problem. We note that in the first term of (\ref{main_result }) the lower integration limit is strictly zero; the associated higher order finite-size effects are accumulated in the last term of (\ref{main_result }). 

A similar expression has been previously given by Sen \cite{Sen}. However the sign of the first leading term of (\ref{integralfinite}) was determined incorrectly; additionally Sen's expression is inadvertently missing a factor of $\pi$ in the denominator of the second term. While the first and second terms of (\ref{main_result }) are dependent on the cutoff, the second term has only a very weak dependence.

The first two terms of (\ref{main_result }) have their origin in the additive Weyl DOS (\ref{ 2dWeylDOs}) while the third cutoff-independent term is additive as well. This is an aspect in which the last term of (\ref{main_result }) denoted $\mathcal{U}_{na}$ is different: it represents a non-additive intrinsic part of the Casimir effect which is sensitive to the topology of the problem. Its functional form is dictated by dimensional analysis: it is of the order $\hbar c$ divided by a macroscopic length scale whose exact form is determined by the details of the problem. If there is only a single length scale $a$, then $\mathcal{U}_{na}\sim\hbar c/ a$; in several cases below we will be able to explain both the sign and magnitude of $\mathcal{U}_{na}$ that are not determined by the dimensional argument. The third and the fourth term of (\ref{main_result }) have the same order of magnitude; in view of their different physical origin we keep these terms separate.

For a circle the effects of curvature and circumference
are not distinguishable. We need a geometry where the two are independent, so that we can 
generalize the result to the case of
an arbitrary boundary. To this end, we treat the wedge-arc
geometry closely related to the sector geometry first considered by Kac \cite{Kac}. 
\begin{figure}
\includegraphics[width=1.0\columnwidth,keepaspectratio]{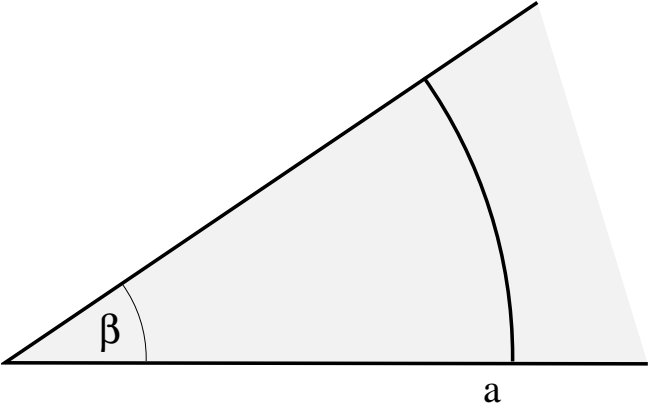} 
\caption{Wedge of opening angle $\beta$ with superimposed arc of radius $a$ in two dimensions.}
\end{figure}

\subsection{Wedge-circular arc geometry in two dimensions; Dirichlet case}

Let us consider a harmonic field theory (\ref{action}) defined inside an infinite wedge of opening angle $\beta$ whose edges enforce the Dirichlet boundary conditions. We are interested in the change in the zero-point energy due to the introduction of a circular Dirichlet arc of radius $a$ and length $s=\beta a$ as shown in Fig. 1. 

The semi-circular ($\beta =\pi$) version of this problem has been studied earlier \cite{Nesterenko2}. Similar to the circular case, a divergence was found which was not removed by the zeta-function regularization method.

The wedge-arc configuration with an arbitrary angle $\beta$ has been
considered in Ref. \cite{Nesterenko3} with the aim of revealing the
regularities in the contributions to
the heat kernel coefficients due to the failure of the boundary to be smooth. The local characteristics of the vacuum
have been discussed by Sakharian and collaborators for a scalar field with Dirichlet boundary
condition in general space-time dimension \cite{Sakharian1} and for the electromagnetic field for $d=3$ \cite{Sakharian2}. 

In polar coordinates $\rho$ and $\varphi$ the boundary-value problem (\ref{Helmholtz}) for this geometry becomes
\begin{eqnarray}
\label{bvcircle}
\left (\frac{1}{\rho}\frac{\partial}{\partial \rho}(\rho \frac{\partial}{\partial \rho}) + \frac{1}{\rho^{2}}\frac{\partial^{2}}{\partial\varphi^{2}} - \frac{\omega^{2}}{c^{2}}\right )u_{\omega} = 0\nonumber\\u_{\omega}(\rho=a,\varphi)= f_{\omega}(\varphi), u_{\omega}(\rho,\varphi=0(\beta))=0
\end{eqnarray}
Seeking the solution in the form $u_{\omega}(\rho, \varphi) = \sum_{n=1}^{\infty}R_{\omega n}(\rho) \sin(\pi n \varphi/\beta)$ and expanding the boundary field into a Fourier series $f_{\omega}(\varphi) = 
\sum_{n
 = 1}^{\infty} f_{\omega n}\sin(\pi n\varphi/\beta)$ gives the boundary-value problem for the radial Fourier coefficients $R_{\omega n}(\rho)$
\begin{equation}
\label{Bessel}
\left (\frac{d^{2}}{d\rho^{2}} + \frac{1}{\rho}\frac{d}{d\rho} - (\frac{\omega^{2}}{c^{2}} + \frac{(\pi n)^{2}}{\beta^{2}\rho^{2}})\right )R_{\omega n} = 0, R_{\omega n}(a)=f_{\omega n}
\end{equation}
The particular solution to (\ref{Bessel}) vanishing at $\rho = 0,\infty$ is 
\begin{eqnarray}
\label{wedge solution}
R_{\omega n}(\rho)& = &f_{\omega n}\frac{I_{\pi n/\beta}(|\omega |\rho/c)}{I_{\pi n/\beta}(|\omega |a/c)} , ~~~~~\rho \leqslant a \nonumber\\
R_{\omega n}(\rho)& = &f_{\omega n}\frac{K_{\pi n/\beta}(|\omega |\rho/c)}{K_{\pi n/\beta}(|\omega |a/c)} , ~~~~\rho > a
\end{eqnarray}
where $I_{\nu}(x)$ and $K_{\nu}(x)$ are modified Bessel functions \cite{AS}. We note that as $n$ increases, the functions (\ref{wedge solution}) become increasingly more localized at the arc. 

If we take the simultaneous $\beta \rightarrow 0$, $a\rightarrow \infty$ limit but keep the length $s = \beta a$ fixed, the geometry of Fig.1 turns into that of an infinite strip of width $s$ whose sides are connected by a straight Dirichlet bridge. Indeed if we substitute $\rho=a+z$ in Eq.(\ref{Bessel}), take the $\beta \rightarrow 0$, $a\rightarrow \infty$, $s = \beta a=const$ limit and employ the fact that the role of the one-dimensional wave vector is played by $q=\pi n/s$, Eq.(\ref{Bessel}) will reduce to Eq.(\ref{bvplane}). The discussion of Section IIIA then implies that in the limit neglecting the curvature of the arc the Casimir energy is given by negative of the half of the zero-point energy of the harmonic field confined to a one-dimensional Dirichlet interval (see Eq.(\ref{surfaceenergysum})). The zero-point energy of the field confined to one-dimensional Dirichlet interval was computed in Section IIIC (see Eq.(\ref{1dzpenergysplit})). With that in mind we infer that neglecting the curvature the Casimir energy is given by 
\begin{equation}
\label{ firstordersplit}
\mathcal{E}^{(1)}=\int_{0}^{\infty~'}\frac{\hbar cq}{2}\left (-\frac{sdq}{2\pi}\right ) + \frac{\pi \hbar c}{48s}
\end{equation}
This illustrates the effect of the geometrical size in general result (\ref{main_result }): the first formally divergent term has its origin in the Weyl DOS given by the first term of (\ref{ 2dWeylDOs}), which is proportional to the arc length. The second intrinsic term of (\ref{ firstordersplit}) has the expected $\hbar c/s$ form and it is not additive. 

In order to account for the effects of curvature we substitute the Fourier representations for $u_{\omega}(\rho,\varphi)$ and $f_{\omega}(\varphi)$ along with the solution (\ref{wedge solution}) in Eq.(\ref{actionsurfaceintegral}). Performing the angular integration we find
\begin{equation}
\label{wedgeaction}
S_{E} = \frac{\hbar}{2T}\sum_{\omega,n}\frac{\beta|f_{\omega n}|^{2}}{2I_{\pi n/\beta}(|\omega|a/c)K_{\pi n/\beta}(|\omega|a/c)}
\end{equation}
where we employed the Wronskian relationship
$ K_{\nu}(z)I_{\nu}^{'}(z) - I_{\nu}(z)K_{\nu}^{'}(z) = 1/z$ \cite{AS}. The action (\ref{wedgeaction}) has the expected form (\ref{form}) and thus the Casimir energy is given by the rule (\ref{ absCasimirenergy})
\begin{equation}
\label{casenergywedge}
\mathcal{E} = \frac{\hbar c}{2\pi a}\sum_{n = 1}^{\infty}~'\int_{0}^{\infty} dx \ln \left (2xI_{\frac{\pi n}{\beta}}(x)K_{\frac{\pi n}{\beta}}(x)\right )
\end{equation} 
In the $\beta \ll 1$ limit this expression can be analyzed with the help of the uniform asymptotic $\nu \gg 1$ expansion of Debye \cite{AS,Schwinger}:
\begin{eqnarray}
\label{Debye}
2\nu(1&+&y^{2})^{1/2}I_{\nu}(\nu y)K_{\nu}(\nu y) = 1 + \frac{1}{8\nu^{2}} \Big\{\frac{1}{1+y^{2}}\nonumber\\& -& \frac{6}{(1+y^{2})^{2}} + \frac{5}{(1+y^{2})^{3}} \Big\} + \mathcal{O}(\frac{1}{\nu^{3}})
\end{eqnarray} 
If only the lowest order term of the Debye expansion is kept, the calculation of the Casimir energy (\ref{casenergywedge}) nearly mirrors that of Section IIIA;  the outcome, in fact, reproduces Eq.(\ref{ firstordersplit}). 

Writing the Casmir energy as $\mathcal{E}=\mathcal{E}^{(1)}+\mathcal{E}^{(2)}+...$, using the next order term of the $\beta \ll 1$ expansion (\ref{Debye}) and integrating we find 
\begin{equation}
\label{ secondorder}
\mathcal{E}^{(2)}=-\frac{\hbar cs}{256\pi a^{2}}\sum_{n=1}^{\infty}~'\frac{1}{n}
\end{equation}
We observe that this term appears to be the length of the arc $s$ multiplied by the square of the curvature $1/a$. Comparing it with the definition of the zeta-function (\ref{zeta}) we see that without the physical cutoff the series in (\ref{ secondorder}) corresponds to the pole of the zeta-function. 

In the macroscopic limit that interests us, the sum in (\ref{ secondorder}) can be transformed into an integral with the help of the definition of Euler's constant $\gamma$ \cite{AS}:
\begin{equation}
\label{Euler_gamma }
\sum_{n=1}^{\infty}~'\frac{1}{n}=\int_{1}^{\infty~'}\frac{dx}{x}+\gamma
\end{equation} 
This brings Eq.(\ref{ secondorder}) into the form
\begin{equation}
\label{ secondordersplit}
\mathcal{E}^{(2)}=\int_{\pi/s}^{\infty~'}\frac{\hbar cq}{2}\left ( -\frac{sdq}{128\pi a^{2}q^{2}}\right )-\frac{\hbar c\gamma s}{256\pi a^{2}}
\end{equation}
where we employed $q=\pi n/s$.
Combining Eqs.(\ref{ firstordersplit}) and (\ref{ secondordersplit}) we arrive at the expression for the Casimir energy of the Dirichlet wedge of arc length $s$ and small opening angle $\beta$ 
\begin{eqnarray}
\label{totenergy}
\mathcal{E}_{Dirichlet}&=& \int_{0}^{\infty~'}\frac{\hbar cq}{2}\left (-\frac{sdq}{2\pi}\right )\nonumber\\&+& \int_{\pi/s}^{\infty~'}\frac{\hbar cq}{2}\left (-\frac{sdq}{128\pi a^{2}q^{2}}\right )\nonumber\\
&-&\frac{\hbar c\gamma s}{256\pi a^{2}}+\frac{\pi \hbar c}{48s}
\end{eqnarray}
The first three terms are the geometrical part, and agrees
with the curvature expansion (\ref{main_result }) which is valid for an
arbitrary curve. Indeed, since Eq.(\ref{totenergy}) is valid in the $\beta =s/a\ll1$ limit, we can take the arc length to be infinitesimally small, $s\rightarrow ds$. Then the first two terms of (\ref{totenergy}) can be viewed as having their origin in the differential DOS
\begin{equation}
\label{ differentialDOS}
dg(q)=-\frac{ds}{2\pi}-\frac{\mathcal{C}^{2}(s)ds}{128\pi q^{2}}
\end{equation}
where $\mathcal{C}=1/a$ is the curvature of the arc. Linearity in $ds$ implies additivity. Then integrating over the boundary we arrive at Eq.(\ref{ 2dWeylDOs}). Similarly, the third term of (\ref{totenergy}) generalizes to the third term of (\ref{main_result }).

The accuracy of the Debye expansion (\ref{Debye}), implies that Eq.(\ref{totenergy}) approximately captures the whole $0\leqslant \beta \leqslant 2\pi$ range. 

\subsection{Wedge-circular arc geometry in two dimensions: periodic boundary conditions}

The effect of topology can be investigated by comparing the
geometry just studied with its periodic counterpart. Consider the wedge-circular arc geometry as in Fig. 1 but assume that the wedge is bounded periodically, rather than 
by a Dirichlet edge, so that the condition $u(\rho, \varphi) = u(\rho, \varphi+\beta)$ applies. The opening
angle will again be assumed to be arbitrary. For $\beta=2\pi$ the 
geometry is that of a plane with a Dirichlet circle; for general $\beta$
we can envision the domain of the field as being the surface of a cone, 
again with a Dirichlet circle at distance $a$ from the apex of the cone \cite{Nesterenko3}.

The Casimir energy of this configuration can be inferred from the solutions of the Dirichlet case. Indeed, the boundary-value problem we need to solve is posed by Eq.(\ref{bvcircle}) with the boundary field satisfying the condition of periodicity, $f_{\omega}(\varphi) = f_{\omega}(\varphi+\beta)$. This implies $u_{\omega}(\rho,\varphi) = \sum_{n=-\infty}^{\infty}R_{\omega n}(\rho) \exp(2\pi i n \varphi/\beta)$. The radial function $R_{\omega n}(\rho)$ satisfies the same Eq.(\ref{Bessel}) with $\pi n/\beta$ being replaced by $2\pi n/\beta$. The Casimir energy can then be inferred from Eq.(\ref{casenergywedge}) as 
\begin{eqnarray}
\label{Casenergyperiodicwedge}
\mathcal{E}& =& \frac{\hbar c}{2\pi a}\int_{0}^{\infty} dx \ln (2xI_{0}(x)K_{0}(x))\nonumber\\
&+& \frac{\hbar c}{\pi a}\sum_{n = 1}^{\infty}~' \int_{0}^{\infty} dx \ln \left (2xI_{\frac{2\pi n}{\beta}}(x)K_{\frac{2\pi n}{\beta}}(x) \right )
\end{eqnarray}
where the first term is due to the fluctuating angle-independent Fourier component of the boundary field $f_{\omega}$ while the sum is a contribution from the angle-dependent components. We note that for $\beta=2\pi$ Eq.(\ref{Casenergyperiodicwedge}) reduces to its circular counterpart \cite{Nesterenko}; this reference, however does not discuss the issue of the physical cutoff. 

The integral entering the first term of (\ref{Casenergyperiodicwedge}) was numerically evaluated by Sen \cite{Sen} and more accurately by Milton and Ng \cite{Ng} with the result
\begin{equation}
\label{adependentuniversal}
\mathcal{E}_{0} = \frac{\hbar c}{2\pi a}\int_{0}^{\infty} dx \ln 2xI_{0}(x)K_{0}(x) = - \frac{0.08808\hbar c}{2\pi a} 
\end{equation} 
This is the intrinsic part of the Casimir effect unique to the periodic geometry; it naturally has no dependence on the opening angle of the wedge. The $\hbar c/a$ form of (\ref{adependentuniversal}) is dictated by dimensional analysis while the negative sign can be understood as due to elimination of the angle-independent modes by the boundary. 

The second term of Eq. (\ref{Casenergyperiodicwedge}) can be analyzed analogously to that of Eq.(\ref{casenergywedge}). Similar to the Dirichlet case studied in Section IVA, the results of the lowest order Debye expansion can be inferred from analysis of periodic one-dimensional interval conducted in Section IIIC. 

The result of the calculation employing next order term of the Debye expansion is nearly identical to (\ref{ secondordersplit}). The only difference is that the lower integration limit in the first term is replaced with $2\pi/s$; this is a consequence of the relation $q=2\pi n/s$ and a sign of sensitivity to topology. Finally the counterpart of Eq.(\ref{totenergy}) has the form
\begin{eqnarray}
\label{totenergyperiodicwedge}
\mathcal{E}_{periodic}&=& \int_{0}^{\infty~'}\frac{\hbar cq}{2}\left (-\frac{sdq}{2\pi}\right )\nonumber\\&+& \int_{2\pi/s}^{\infty~'}\frac{\hbar cq}{2}\left (-\frac{sdq}{128\pi a^{2}q^{2}}\right )\nonumber\\
&-&\frac{\hbar c\gamma s}{256\pi a^{2}}+\frac{\pi \hbar c}{12s}-\frac{0.0880\hbar c}{2\pi a}
\end{eqnarray}
This expression is again in agreement with general results, Eqs. (\ref{main_result })-(\ref{integralfinite}), and the differences between the Dirichlet (\ref{totenergy}) and periodic (\ref{totenergyperiodicwedge}) Casimir energies are due to different topology. The first source of the difference lies in different non-additive pieces of the effect $\mathcal{U}_{na}$ (see Eq.(\ref{main_result })). The second source, unique to two-dimensional geometry, is due to the second sub-leading term of the Weyl DOS (\ref{ 2dWeylDOs}). 

\subsection{Application:  which geometry has lower energy - Dirichlet or periodic ?}

Here we apply our general results to the situation where the physics is dominated by the cutoff-independent part of the Casimir effect. Earlier we demonstrated that the Casimir energy due to a smooth Dirichlet curve is generally cutoff-dependent and dominated by the geometrical contribution (\ref{integralfinite}) generated by the Weyl DOS (\ref{ 2dWeylDOs}). However in the difference in energy between geometries that are only topologically distinct, only the cutoff-independent parts contribute. Indeed in the macroscopic limit $\omega_{0}s/c\gg 1$ the difference between (\ref{totenergyperiodicwedge}) and (\ref{totenergy}) is given by
\begin{equation}
\label{difference }
\triangle\mathcal{E}=\mathcal{E}_{periodic}-\mathcal{E}_{Dirichlet}=\frac{\pi \hbar c}{16s}-\frac{0.08808\hbar c}{2\pi a}+\frac{\hbar cs\ln2}{256\pi a^{2}}
\end{equation}
where the last term originates from the second term of the Weyl DOS (\ref{ 2dWeylDOs}). It is straightforward to realize that for any relationship between $a$ and $s$ the periodic geometry has the larger energy, $\triangle\mathcal{E}>0$. In order to see this we minimize (\ref{difference }) with respect to $a$, substitute the outcome back in (\ref{difference }) and observe that for any $s$ we have $\triangle\mathcal{E}>0$. 

\section{Summary}

The intrinsic part of the Casimir energy is interesting because it is cutoff-independent and thus does not depend on the material properties of the boundary. However, the cutoff-dependent terms have physical meaning and would contribute to the Casimir stress on a sphere (for example), since a virtual change in the radius of the sphere would change the cutoff-dependent contributions to the energy. 

We have shown how the Weyl expansion determines the cutoff-dependent terms. In most cases there is a clear separation of the Weyl and intrinsic contributions into the energy as given by Eq.(\ref{splitzpenergy }). Then the cutoff-dependent contributions to the energy depend on a small set of coefficients characterizing the sensitivity to area and curvature of the surface; with these in hand the cutoff-dependent contributions to the Casimer energy are determined by geometry. 

This fails in even dimensions, because the last relevant term of the Weyl DOS expansion is proportional to $1/q^{2}$. The corresponding contribution to the Casimir energy is only logarithmically dependent on the cutoff frequency. Then the separation of the cutoff-dependent and intrinsic parts of the Casimir energy is no longer clear cut. Here the only guaranteed way to arrive at physically measurable answer is to use a cutoff method that emulates the transmission properties of the boundaries. Following this path we demonstrated that in the macroscopic limit the Casimir energy due to an arbitrary Dirichlet boundary in two-dimensions is given by Eq.(\ref{main_result }). Here the concept of the Weyl DOS again plays a prominent role but separation of the intrinsic part of the effect requires careful analysis. This explains why the zeta-function regularization method does not work for circular \cite{Romeo,Nesterenko} and semi-circular \cite{Nesterenko2} boundaries in two dimensions. Although we only analyzed the two-dimensional case, we expect that the mystery of divergent Casimir self-stress in general even space dimension \cite{Bender} is solved similarly. 

The approach to the Casimir effect employed in this work focused on the energy content of field modes eliminated by the Dirichlet boundary. The applicability of this method is not limited to the scalar field theory (\ref{action}), Dirichlet boundaries, or zero-temperature limit. One of the additional benefits of our method is the ability to provide the physical insight necessary to predict the sign of the Casimir self-stress. 

\section{Acknowledgments}

We are grateful to M. Schaden for a stimulating conversation that greatly informed our research. This work was supported by the Thomas F. Jeffress and Kate Miller Jeffress Memorial Trust. A portion of this work was performed when the first author enjoyed the hospitality of the Kavli Institute for Theoretical Physics in Santa Barbara as a participant of the 2008 program \textit{The theory and practice of fluctuation induced interactions} where this research was supported by the National Science Foundation under Grant No. PHY05-51164. We thank the participants of this program for their interest in this work and especially T. Emig, J. Feinberg, M. Kardar, I. Klich and R. Podgornik for discussions. We also thank A. A. Saharian, M. Bordag and D. V. Vassilevich for providing us with valuable references.


\begin{thebibliography}{10}

\bibitem{Casimir} H. B. G. Casimir, \textit{Proc. K. Ned. Akad. Wet.} \textbf{51}, 793 (1948).

\bibitem{experiment} M. J. Sparnaay, Physica (Amsterdam) \textbf{24}, 751 (1958); S. K. Lamoreaux, Phys. Rev. Lett. \textbf{78}, 5 (1997); U. Mohideen and A. Roy, Phys. Rev. Lett. \textbf{81}, 4549 (1998);G. Bressi, G. Carugno, R. Onofrio, and G. Ruoso, Phys. Rev. Lett. \textbf{88}, 041804 (2002).

\bibitem{CasReviews} For a review of the state of theoretical and experimental efforts see K. A. Milton, J. Phys. A: Math. Gen. \textbf{37}, R209 (2004) and G. L. Klimchitskaya and V.M. Mostepanenko, Contemp. Phys. \textbf{47}, 131 (2006), respectively.

\bibitem{Power} E. A. Power, \textit{Introductory Quantum Electrodynamics} (American Elsevier Publishing Company, Inc., New York, 1965), Section 3.4 and Appendix 1. 

\bibitem{zeta} J. S. Dowker and R. Critchley, Phys. Rev. D \textbf{13}, 3224 (1976).

\bibitem{dim} J. Ambj\o rn and S. Wolfram, Ann. Phys. \textbf{147}, 1 (1983), and references therein.

\bibitem{Bender} C. M. Bender and K. A. Milton, Phys. Rev. D \textbf{50}, 6547 (1994).

\bibitem{Milton} K. A. Milton, \textit{The Casimir Effect: Physical Manifestations of Zero-Point Energy}, (World Scientific, 2005), and references therein. 

\bibitem{Sen} S. Sen, J. Math. Phys. \textbf{22}, 2968 (1981); Phys. Rev. D \textbf{24}, 869 (1981).

\bibitem{DK} J. S. Dowker and G. Kennedy, J. Phys. A: Math. Gen. \textbf{11}, 895 (1978).

\bibitem{CD} D. Deutsch and P. Candelas, Phys. Rev. D \textbf{20}, 3063 (1979); P. Candelas, Ann. Phys., NY \textbf{143}, 241(1982); P. Candelas, Ann. Phys., NY \textbf{167}, 257 (1986).

\bibitem{Weyl} H. P. Baltes and E.R. Hilf, \textit{A Review of Weyl`s Problem: The Eigenvalue Distribution of the Wave Equation for Finite Domains and its Applications in the Physics of Small Systems} (Bibliographisches Institut, Zurich, 1976), and references therein; D. V. Vassilevich, Phys. Rep. \textbf{388}, 279 (2003), and references therein;  M. Brack and R. K. Bhaduri, \textit{Semiclassical Physics} (Westview Press, 2008), Chapter 4, and references therein; M. Bordag, G. L. Klimchitskaya, U. Mohideen,  and V. M. Mostepanenko, \textit{Advances in the Casimir Effect} (Oxford University Press, 2009), Chapter 4, and references therein.

\bibitem{KK} O. Kenneth and I. Klich, Phys. Rev. Lett. \textbf{97}, 160401 (2006).

\bibitem{Schaden} M. Schaden, Phys. Rev. A \textbf{73}, 042102 (2006), and references therein.

\bibitem{Matsubara} T. Matsubara, Prog. Theor. Phys. \textbf{14}, 351 (1955).

\bibitem{Kogut} J. B. Kogut, Rev. Mod. Phys. \textbf{51}, 659 (1979), and references therein.

\bibitem{Kleinert} H. Kleinert, \textit{Path Integrals in Quntum Mechanics, Statistics, Polymer Physics and Financial Markets}, Fourth Edition (World Scientific Publishing Company, 2006), Chapter 2. 

\bibitem{Brevik}  I. Brevik and E. Elizalde, Phys. Rev. D \textbf{59}, 5319 (1994).

\bibitem{contour} N.G. Van Kampen, B.R.A. Nijboer, and K. Schram, Phys. Lett. A \textbf{26}, 307 (1968).

\bibitem{Jaffe} N. Graham, R. L. Jaffe, V. Khemani, M. Quandt, M. Scandurra and H. Weigel, Nucl. Phys. B \textbf{645}, 49 (2002); Phys. Lett. B \textbf{572}, 196 (2003); N. Graham, R. L. Jaffe and H. Weigel, Int. J. Mod. Phys. A \textbf{17}, 846 (2002); N. Graham and K. D. Olum, Phys. Rev. D \textbf{67}, 085014 (2003); Erratum-ibid. D \textbf{69}, 10990 (2004); K. D. Olum and N. Graham, Phys. Lett. B \textbf{554}, 175 (2003); R. L. Jaffe, arXiv:hep-th/0307014v2; N. Graham, R. L. Jaffe, V. Khemani, M. Quandt, O. Schroeder and H.Weigel, Nucl. Phys. B \textbf{677}, 379 (2004).

\bibitem{Barton} G. Barton, J. Phys. A: Math. Gen. \textbf{37}, 1011 (2004).

\bibitem{critique} The conclusions of Graham, Jaffe and co-workers \cite{Jaffe} were criticized in K. A. Milton, Phys. Rev. D \textbf{68}, 065020 (2003) and J. Phys. A: Math. Gen. \textbf{37}, 6391(2004); further discussion of surface divergences can be found in Milton's review \cite{CasReviews}, in S. A. Fulling, J. Phys. A: Math. Gen. \textbf{36}, 6857 (2003) and in M.Bordag and D.V.Vassilevich, Phys. Rev.D \textbf {70} 045003 (2004).

\bibitem{BD} R. Balian and B. Duplantier, Ann. Phys. (N. Y.) \textbf{112}, 165 (1978).

\bibitem{Kac} M. Kac, \textit{The American Mathematical Monthly} \textbf{73}, No. 4, Part 2: Papers in Analysis, 1 (1966), and references therein.

\bibitem{Brownell}  F. H. Brownell, J. Math. Mech, \textbf{6}, 119 (1957); see also V. Ya. Ivrii, Funktsional. Anal. i Prilozhen. \textbf{14}, 25 (1980) (in Russian) [Funct. Anal. Appl. \textbf{14}, 98 (1980)] and R. B. Melrose, Proc. Sympos. Pure Math., Univ. Hawaii, Honolulu, Hawaii,
1979, Proc. Sympos. Pure Math., XXXVI, Amer. Math. Soc., Providence, R.I., 257 (1980).

\bibitem{integral}  E. T. Whittaker and G. N. Watson,  \textit{A Course in Modern Analysis}, fourth edition (Cambridge University Press, 1990), Section 7.21;  see also L. D. Landau and E. M. Lifshitz, \textit{Statistical Physics}, vol.V, Part 1, third edition, revised and enlarged by E. M. Lifshitz and L. P. Pitaevskii (Pergamon, 1980), Section 59, Eq.(59.10a) where the Euler-Maclaurin formula was employed in a related context.

\bibitem{conformal} K. Johnson, Acta Physica Polonica B \textbf{6}, 865 (1975); M. L\"uscher, K. Symanzik, and P. Weisz, Nucl. Phys. B \textbf{173}, 365 (1980); M. L\"uscher, Nucl. Phys. B \textbf(173), 317 (1981); H. W. Bl\"ote, J. L. Cardy , and M. P. Nightingale, Phys. Rev. Lett. \textbf{56}, 742 (1986); I. Affleck, Phys. Rev. Lett. \textbf{56}, 746 (1986).

\bibitem{Elizalde}  E. Elizalde,  J. Phys. A: Math. Gen. \textbf{36},  L567 (2003), \textit{ibid.}  \textbf{39}, 6299 (2006).

\bibitem{Boyer0} T. H. Boyer, Phys. Rev. \textbf{174}, 1764 (1968).

\bibitem{geometry} K. Stewartson and R. T. Waechter, Proc. Cambridge Philos. Soc. \textbf{69}, 353 (1971); see also A. Pleijel, Arkiv. Matematik \textbf{2}, 553 (1954).

\bibitem{Nesterenko2} V. V. Nesterenko, G. Lambiase, G. Scarpetta, J. Math. Phys. \textbf{42}, 1974 (2001), Section V.

\bibitem{Nesterenko3} V. V. Nesterenko, I. G. Pirozhenko, and J. Dittrich, Class. Quantum Grav. \textbf{20}, 431 (2003).

\bibitem{Sakharian1} A.A. Saharian, A.S. Tarloyan, J. Phys. A: Math. Gen. \textbf{38}, 8763 (2005).

\bibitem{Sakharian2} A. A. Saharian, Eur. Phys. J. C \textbf{52}, 721 (2007).

\bibitem{AS} \textit{Handbook of Mathematical Functions}, edited by M. Abramowitz and I. A. Stegun (Dover, New York, 1972).

\bibitem{Schwinger} K. A. Milton, L. L. DeRaad, Jr. and J. Schwinger, Ann. Phys. (N.Y.) \textbf{115}, 388 (1978).

\bibitem{Nesterenko} V. V. Nesterenko and I. G. Pirozhenko, J. Math. Phys. \textbf{41}, 4521 (2000).

\bibitem{Ng} K. A. Milton and Y. J. Ng, Phys. Rev. D \textbf{46}, 842 (1992).

\bibitem{Romeo} S. Leseduarte and A. Romeo, Ann. Phys. \textbf{250}, 448 (1996); G. Cognola, E. Elizalde and K. Kirsten, J. Phys. A: Math. Gen. \textbf{34}, 7311 (2001).

\end{thebibliography}
\end{document}